\documentclass[10pt]{iopart}
\usepackage{amssymb}

\newcommand{\be}{\begin{equation}} 
\newcommand{\ee}{\end{equation}} 
\newcommand{\ds}{\displaystyle}
\usepackage{epsfig}
\begin{document}

\title[Detector optimisation]{VIRGO detector optimisation for gravitational waves by inspiralling binaries}

\author{Alessandro D.A.M. Spallicci$^1$, Sofiane Aoudia$^{1,2}$,
Jos\'e de Freitas Pacheco$^1$, Tania Regimbau$^1$ and Giorgio Frossati$^3$}

\address{{ $^1$ UMR 6162 D\'ept. d'Astrophysique Relativiste ARTEMIS, CNRS Observatoire de la C\^{o}te d'Azur,  France\footnote{BP 4229, Bd de l'Observatoire, 06304 Nice}}\\
{$^2$ Universit\'e de Nice Sophia Antipolis, France }\\
{$^3$ Kamerlingh Onnes Laboratorium, Rijksuniversiteit van Leiden, Nederland}}

\bigskip

\noindent{\small Emails: spallicci@obs-nice.fr, aoudia@obs-nice.fr, pacheco@obs-nice.fr}

{\small $~~~~~$regimbau@obs-nice.fr, frossati@physics.leidenuniv.nl}

\bigskip


\begin{abstract}
For future configurations, we study the relation between the abatement of the noise sources and the Signal to Noise Ratio (SNR) for coalescing binaries.  
Our aim is not the proposition of a new design, but an indication of where in the bandwidth or for which noise source, a noise reduction would be most efficient. We take VIRGO as the reference for our considerations, solely applicable to the inspiralling phase of a coalescing binary. Thus, only neutron stars and small black holes of few solar masses are encompassed by our analysis. The contributions to the SNR given by final merge and quasi-normal ringing are neglected. 
It is identified that i) the reduction in the mirror thermal noise band provides the highest gain for the SNR, when 
the VIRGO bandwidth is divided according to the dominant noises; ii) 
it exists a specific frequency at which lies the potential largest increment in the SNR, and that the enlargement of the bandwidth, where the noise is reduced, produces a shift of such optimal frequency to higher values; iii) the abatement of the pendulum thermal noise provides the largest, but modest, gain, when noise sources are considered separately. Our recent astrophysical analysis on event rates for neutron stars leads to a detection rate of one every 148 or 125 years for VIRGO and LIGO, respectively, while a recently proposed and improved, but still conservative, VIRGO configuration would provide an increase to 1.5 events per year. Instead, a bi-monthly event rate, similar to advanced LIGO, requires a 16 times gain. We analyse the 3D (pendulum, mirror, shot noises) parameter space showing how such gain could be achieved.
\end{abstract}

\section{Introduction}

Improved detectors are currently under consideration in Europe. The motivation of this work is the analysis of the cost/benefit ratio in support to the forthcoming R\&D effort, a major concern being how to gear efforts such that the improvements will sensibly increase the probability of detection and give birth to an observational relativistic astrophysics. 
To such end, this work takes a very pragmatic approach. The design of future interferometers is well beyond the scope of this paper and instead we deal with the detector from an user view point. Once the most efficient utilisation strategy shall be identified, detailed design and R\&D should follow accordingly. In this respect, the approach adopted herein proposes the end-user, i.e. the astrophysicist, to guide the requirements on the design of the interferometers. A not very different approach was taken by Babusci and Giovannini (2001) 
who investigated stochastic and isotropic gravitational wave backgrounds produced in pre-big-bang models. \\
Scientific interest, knowledge of the model, and number of events hint to coalescing binaries as primary candidates. Thus, improvements in signal to noise ratio (SNR) by noise abatement, for inspiralling binaries of small chirp mass, are discussed. For this exercise, we make reference to the VIRGO interferometer.\\
The structure of the paper is the following. 
In the next section, we show the event rates out of a recent analysis (de Freitas Pacheco et al., 2004). In the third section, we briefly summarise the main expressions that determine the SNR for inspiralling binaries and present the main features of  the VIRGO sensitivity curve. In the fourth section, we perform the analysis leading to our main findings on the relation between noise abatement and the increment in the SNR. In the fifth section, we estimate the detection rate of a recently proposed VIRGO configuration and address the necessary improvements to acquire a detection rate similar to advanced LIGO.     

\section{Event rates}

The estimates of the coalescence rate of NS-NS binaries are performed in two main steps. Firstly, the merging rate
in our Galaxy is evaluated and then, by assuming that this value is typical, extrapolations to the local universe
are performed under the assumption of some adequate scaling. The galactic rate has been estimated by different 
authors and ranges from $10^{-6}/yr^{-1}$ up to few times $10^{-5}/yr^{-1}$, but values as high as $(2-3)\times
10^{-4}/yr^{-1}$ have already been reported (Tutukov \& Yungelson 1993; Lipunov 1997; Kalogera et al. 2004). Here
we adopt the approach by de Freitas Pacheco (1997) and de Freitas Pacheco et al. (2004).

Let us suppose a massive binary (masses of the components higher than 9 M$_{\odot}$) formed at the
instant $t'$. Let $\tau_*$ be the mean evolutionary timescale for the system to evolve into two
neutron stars, typically of the order of $10^8$ yr. Define $P(\tau)$ as the probability per unit of time
for a newly formed NS-NS binary to coalesce in a timescale $\tau$ and define $R_*(t)$ as the star
formation rate (in M$_{\odot}yr^{-1}$). Then, the galactic coalescence rate at instant $t$ is 

\begin{equation}
\nu_c(t) = f_b\beta_{ns}\lambda\int^{(t-\tau_*-\tau_0)}_{\tau_0}P(\tau)R_*(t-\tau_*-\tau_0)d\tau
\end{equation} 
where $f_b$ is the fraction of massive binaries formed among all stars, $\beta_{ns}$ is the fraction
of formed binaries which remain bounded after the second supernova event and $\lambda$ is the
fraction of formed stars in the mass interval 9-40 M$_{\odot}$. We have set $\tau_0$ as the minimum
timescale for a NS-NS binary to coalesce. Numerical simulations discussed in de Freitas Pacheco
et al. (2004) allow the determination of $\tau_0$, the product $f_b\beta_{ns}$ and the 
probability distribution $P(\tau)$. One obtains $\tau_0 = 2\times 10^5$ yr, $P(\tau) \propto 1/\tau$
and $f_b\beta_{ns} = 0.00326$. This latter value corresponds to a natal kick velocity dispersion
of about 80 km/s. If we adopt the galactic star formation history derived by Rocha-Pinto et al.
(2000), it results for the present galactic NS-NS coalescence rate $\nu_S = (1.7\pm 1.0)\times 10^{-5}\,
yr^{-1}$, where the estimated uncertainty is essentially due to uncertainties in the natal kick
velocity dispersion and in the actual ratio between the number of single to binary pulsars.
A similar calculation for a typical elliptical galaxy with an absolute
magnitude $M_B = -20.7$ gives a coalescence rate $\nu_E = 8.6\times 10^{-5}\, yr^{-1}$ (de Freitas
Pacheco et al. 2004). The weighted local coalescence rate was obtained assuming a fraction of
65\% of spirals and 35\% of ellipticals and lenticulars (S0's) and an elliptical-to-spiral
luminosity ratio of 1.26. The resulting rate is $\nu_c = 3.4\times 10^{-5}\, yr^{-1}$.

Extrapolation from the local coalescence rate to the expected rate within a spherical
volume of radius D probed by the detector is made through the scale factor $K_B(D)$. This
factor is defined as the ratio between the total blue luminosity within the considered volume
and the Milk Way luminosity. We use here the scale factor computed by de Freitas Pacheco et al.
(2004), who considered the distribution of galaxies derived from the LEDA databasis and
included the contribution of the Great Attractor, centered at the Norma cluster.
The expected rates as a function of distance are given in {\it tab. 1}.

\vskip 10pt
\centerline{
\begin{tabular}{||c|c||} \hline \hline
{\bf D (Mpc)} & {\bf Event rate }~~$yr^{-1}$ \\ \hline
10& $3.44\times 10^{-3}$\\ \hline
15& $2.00\times 10^{-2}$\\ \hline
20& $3.00\times 10^{-2}$\\ \hline
30& $6.44\times 10^{-2}$\\ \hline
40& $1.21\times 10^{-1}$\\ \hline
50& $2.20\times 10^{-1}$\\ \hline
70& $7.08\times 10^{-1}$\\ \hline
100& 1.63\\ \hline
150& 2.30\\ \hline
200& 5.44\\ \hline
300& 18.4\\ \hline
\end{tabular}}
\vskip 3pt
\centerline{
{\small\it Tab. 1 Expected event rates as a function of distance.}}

\section{Inspiralling binaries and VIRGO features}

Matched filtering is the baseline detection tool, although computationally demanding. It enhances the signal by the square root of the number of observed cycles in the binary.
For our analysis, the simplest post-Newtonian model is assumed. A more accurate model may marginally change the absolute values of the SNR but not their relative values we are interested in. A simplified expression, derived by the Peters-Mathews model (Peters, 1964; and Mathews, 1963) is given by (Spallicci et al., 1997):

\begin{equation}
\frac{S}{N} = 2 \sigma \tilde{A}\sqrt{\int_0^\infty \frac{1}{f^{7/3}I(f)}df}
\end{equation}

where $I(f)$ is the one-sided spectral density of the noise,

\begin{equation}
\tilde{A} = \sqrt{\frac{5}{24}}\frac{1}{\pi^{2/3}}\frac{{\cal M}^{5/6}}{r}
~~~~~~~~~~~
{\cal M} = (\mu M^{2/3})^{3/5} = \frac{(M_1 M_2)^{3/5}}{(M_1 + M_2)^{1/5}}
\end{equation}
${\cal M}, \mu, M, M_1, M_2$ being the chirp, reduced, total and single masses of the 
binary. The coefficient $\sigma = 2/5$ takes into account the average relative orientation between binary
and detector (the discrepancy with more accurate expressions is around 5\%).
The VIRGO sensitivity curve, {\it fig. 1}, is given by\footnote{http://www.virgo.infn.it}:

\begin{equation}
\fl{
I_{old}(f) = 3.8 \times 10^{-51}f^2 + 9.5 \times 10^{-46} +
1.6 \times 10^{-43}f^{-1} + 8.3 \times 10^{-37}f^{-5} 
}
\end{equation}
and it is divided into the following zones dominated by a specific noise:
a. 1 - 2.3   Hz seismic, b. 2.3 - 52 Hz thermal pendulum, c. 52 - 148 Hz thermal mirror, 
d. 148 - 10.000 Hz shot. 
We define (Spallicci, 2003) the gain simply as (the upper limit of the integrals is known to be irrelevant 
if sufficiently large to encompass the total inspiral):

\begin{equation}
G = \frac{(SNR)_{new}}{(SNR)_{old}} = 
\left (
\frac
{\ds \int_{1~Hz}^{10~KHz}\frac{df}{I_{new}f^{7/3}}}
{\ds \int_{1~Hz}^{10~KHz}\frac{df}{I_{old}f^{7/3}}} 
\right )^{1/2}
\label{eq:gspot}
\end{equation}
where $I_{new}(f)$ generally indicates an improved spectral noise density.  

\section{Analysis for an improved SNR} 

We emphasize that 
eq. (\ref{eq:gspot}) is a ratio and that most of the consequences of simplifying assumptions (on the antenna and the source) will be
not influential.
Our considerations are applicable solely to the inspiralling phase of a coalescing binary. This is equivalent to say that only neutron stars and small black holes of few solar masses are encompassed by this analysis. The contributions to the SNR given by final merge and quasi-normal ringing are neglected. \\ 
We introduce the noise abatement factor $k_f$ for which:

\begin{equation}
\fl{~~~~~~~~~~~~
\frac{I_{new}}{I_{old}} = \frac{1}{k_f^2}\left[u\left(f + F\right ) - 
u\left(f - F \right )\right ] 
+ \left[1 - u\left(f + F\right ) + 
u\left(f - F\right )\right ]
}
\end{equation} 
where 
\[
F = \frac{\Delta f}{2} - f_0~~~~~u(f) = 0 (f < 0), 0.5 (f = 0) , 1 (f > 0)
\]
We note that the Gate function is built on the Heaviside Step $u(f)$ function.
The operator $k_f$ acts on a sub-band $\Delta f$ centered on $f_0$. Therefore, the function $I_{new}$ takes the values $I_{old}$ or $I_{old}/k_f^2$, outside and within the chosen window, respectively. \\
The first investigation aims to identify the SNR improvement for abatement of the noise in one of the four a-d sub-bands defined in the previous section. We improve each of the four sub-bands in turns; {\it fig. 2} shows the behavior of the SNR gain for different values of noise abatement.     
We observe that the abatement in the thermal mirror noise band (c) provides the largest SNR gain. \\
The second investigation aims to identify the optimal frequency around which the noise abatement should be performed. We firstly look for the single frequency at which the noise reduction is more beneficial, identified as 52 Hz ({\it fig. 3}). Further, we analyse where the noise reduction in a given bandwidth of $10$ or $100$ Hz would be more beneficial ({\it 
fig. 3}). The increment of the bandwidth shifts the optimal frequency to larger values: for a 100 Hz bandwidth, the optimal frequency is 75 Hz. Furthermore, equivalent situations achieve the same gain in the SNR offering alternative strategies. {\it Fig. 3} reports a similar gain G = 2.5 for $k_f = 3$ at 75 Hz in a bandwidth of 100 Hz or for 
$k_f = 10$ at 30 Hz in a bandwidth of 10 Hz. The maximum gain is G = 8 for a bandwidth of 100 Hz centred at 75 Hz. \\
In the 52-148 Hz band, the major limiting factor is coming from the
thermal noise of the mirrors but a large noise reduction would imply
lowering the contributions from the pendulum and the shot noises too, since a large
reduction in thermal noise is useless without also reducing the other noise sources. \\
If instead, we consider the noise abatement per noise source, the new spectral noise density is given by:

\begin{equation}
I_{new} = \sum_{j = 1}^{r}h_{j}^2 + \left (\frac{h_{r + 1}}{k_n}\right )^2
\end{equation} 
having attributed a noise reduction factor $k_n$ to a single source, out of $r + 1$ noises. 
Indeed, the sole reduction of the thermal mirror noise leads to a gain G = 1.35, {\it fig. 4}.  
When we consider the four major noise sources and apply to each of them, in turns, a reduction factor $k_n$, {\it fig. 4}, the gain G is never larger than 1.55 (thermal pendulum noise). Obviously, asymptotic 
gain values are achieved, whenever only one single noise source is dealt with, while the others are left unchanged. 

\section{Detection rates}
The actual specifications on the VIRGO sensitivity show that a SNR of 7 is achieved at 13 Mpc for neutron stars binaries
combined with a false alarm rate (FAR) of 1 per year. This distance corresponds to an event rate of 1 every 148 years ({\it tab. 1}), while the event rate for LIGO is 1 each 125 years (Pacheco et al., 2004), same SNR and FAR being at 14 Mpc. A gain G of about 8 in the SNR, {\it fig. 3} would allow VIRGO to explore the universe up to about 100 Mpc with the same SNR and FAR. To the latter values, a detection rate of 1.5 event per year is associated.\\
Recently, an improved configuration for VIRGO has been proposed (Punturo, 2004), its sensitivity curve being shown in 
{\it fig. 5}. The new sensitivity is obtained by reducing the pendulum noise by a factor 28, the thermal noise by 7 and the shot noise by 4, throughout the spectrum. The associated gain G is 7.5. The Punturo (2004) configuration is rather conservative and is currently analysed for further improvement. Future configurations will likely consider  
a cryogenic design\footnote{
The suspended test masses thermal noise power spectrum depends upon the temperature, the strain energy and the loss angle of the mirror (Harry et al., 2002; Bondu et al., 1998).
As demonstrated by Braginsky et al. (1998, 2003), the thermo-elastic effect is present both in the mirror and in the coating (thermo-dynamical contribution). 
The direct temperature dependence of the fluctuation power spectrum suggests that, if we want gain an order of magnitude in terms of the space-time strain $h$, we must decrease the temperature of about two orders of magnitude, that is go to cryogenic temperatures. Planning entails use of crystalline Silicon to realize the suspension for its lower loss angle than fused silica; its thermal expansion coefficient goes essentially to zero before 10 K. Also, the thermal conductivity of the Silicon increases at low temperature pushing up in frequency the residual thermo-elastic peak. The latter property of Si is also fundamental for its use in an advanced interferometric detector, 
in which the power circulating is so high that the thermal lensing effect in the mirrors could change the optical properties of the Fabry-Perot cavities, affecting the whole functioning. The Silicon thermo-mechanical properties suggest it as a good candidate also for the mirrors. All the thermo-dynamical phenomena related to the substrate would disappear using Silicon mirrors at cryogenic temperatures. 
Different is the behavior of the coatings. Recent preliminary measurements  showed that the loss angle of Ta$_2$O$_5$ coatings doesn't decrease with the temperature. In any case, at least a reduction of a factor 10 due to the direct temperature dependence is expected in terms of noise amplitude spectral density. 
Obviously, in order to use Silicon mirrors, it is necessary to change the wavelength of the laser. In fact, Si transmittivity is very poor at the usual $1.064 \mu m$ wavelength used in the current Nd:YAG laser in GW experiments, but it increases greatly  at $1.5 \mu m$. }.
From {\it tab. 1}, we evince that it is 
mandatory that a detector reaches at least 80 Mpc to get an yearly rate of events.
A significant further relevant improvement imposes VIRGO to go beyond 210 Mpc for one event each two months, as advanced LIGO.        
{\it Fig. 6} shows which combinations of concurring abatements on the three most dominant noises (thermal pendulum, thermal mirror, shot) produces a gain G = 16, necessary to achieve 210 Mpc with SNR = 7 with an yearly FAR. 
The dots represent integer values of the noise reduction factors for thermal pendulum and mirror, joined by isolines of 
the shot noise reduction factor. 
     
\section{Conclusions}

We have analysed the efficiency of potential improvements on the actual VIRGO configuration versus the SNR for coalescing binaries.  
Recently proposed configurations would explore a sphere of less than 100 Mpc of radius (SNR = 7, yearly FAR) and provide 1.5 detections per year, while the actual configuration would detect 1 event each 148 years, exploring up to 13 Mpc (SNR = 7, yearly FAR). A bi-monthly rate requires an improvement of a factor 16 in gain. The noise reduction parameter space necessary to achieve such gain has been explored.

\section{Acknowledgments}

This work was supported by the European Space Agency with a G. Colombo Senior Research Fellowship to A. Spallicci. A. Brillet, E. Chassande-Mottin, M. Punturo, and J.-Y. Vinet are acknowledged for discussions. 

\section*{References}
\noindent
{Babusci D. and Giovannini M., 2001. Int. J. Mod. Phys. D, {\bf 10}, 477.}
\\ 
Bondu F., Hello P. and Vinet J.-Y., 1998. VIR-NOT-LAS-1390-108.
\\
Braginsky V.B., Gorodetsky M.L. and Vyatchanin S.P., 1999. Phys. Lett. A, {\bf 264}, 1.
\\
Braginsky V.B. and Vyatchanin S.P., 2003. Phys. Lett. A, {\bf 312}, 244.
\\
de Freitas Pacheco J.A., 1997. Astrop. Phys., {\bf 8}, 21.
\\
de Freitas Pacheco, Regimbau T., Vincent S. and Spallicci A., 2004, in submission.
\\
Harry G.M., Gretarsson A.M., Saulson P.R., Kittelberger S.E., Penn S.D., Startin W.J., Rowan S., Fejer M.M., Crooks D.R.M., Cagnoli G., Hough J. and Nagawaka N., 2002. Class. Quant. Grav., {\bf 19}, 897. 
\\
Kalogera V.,  Kim C., Lorimer D.R., Burgay M., D'Amico N., Possenti A., Manchester R.N., Lyne A.G., Joshi B.C., McLaughlin M.A., Kramer M., Sarkisian J.M. and Camilo F., 2004. Astrophys. J., {\bf 601}, L179.  
\\
Lipunov V.M., Postnov K.A. and Prokhorov M.E., 1997. Mon. Not. R. Astron. Soc., {\bf 288 }, 245. 
\\
Peters P.C., 1964. Phys. Rev., {\bf 136}, 1124.
\\
Peters P.C. and Mathews J., 1963. Phys. Rev., {\bf 131}, 435.
\\
Punturo M., 2004. VIR-NOT-PER-1390-284.
\\
Rocha-Pinto H.J., Scalo J., Maciel W.J. and Flynn C., 2000. AP. J., {\bf 531}, L115.
\\
Schutz B.F., 1986. Nature, {\bf 323}, 310.
\\
Spallicci A., 2003. VIR-NOT-OCA-1390-257.
\\
Spallicci A., Kr\'olak A. and Frossati G., 1997. Class. Quantum Grav., {\bf 14}, 577.
\\
Tutukov A.V. and Yungelson L.R., 1993. Mon. Not. R. Astron. Soc., {\bf 260}, 675.

\begin{figure}[tbp]
\begin{center}
\epsfig{file=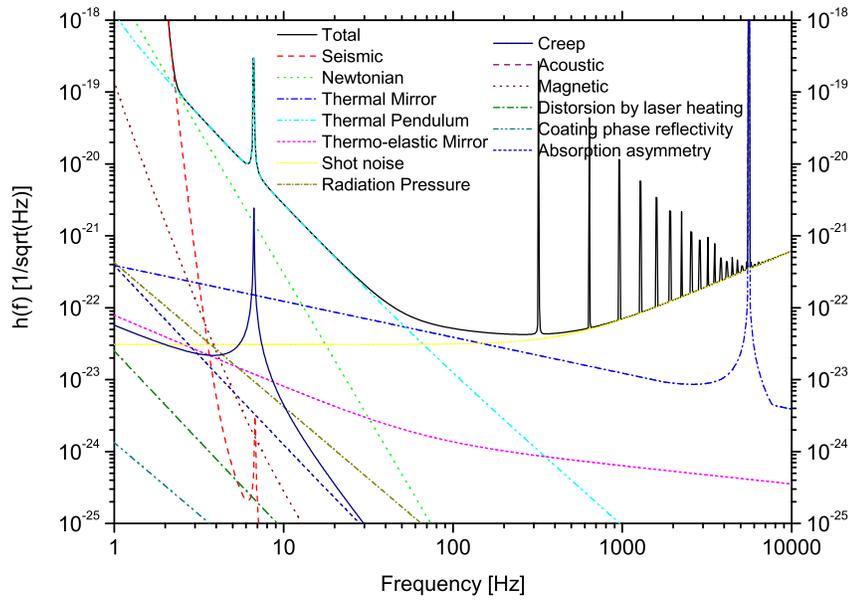}
\end{center}
\caption{VIRGO strain sensitivity $h(f)/\sqrt{Hz}$.}
\label{fig:fig1}
\end{figure}

\begin{figure}[tbp]
\begin{center}
\epsfig{file=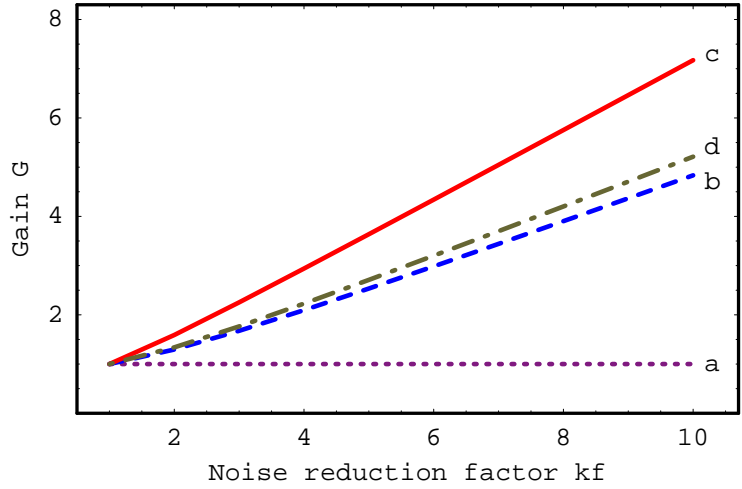}
\end{center}
\caption{Gain G versus noise reduction factor $k_f$ for the four sub-bands a-d (a. 1 - 2.3 Hz seismic, b. 2.3 - 52 Hz thermal pendulum, c. 52 - 148 Hz thermal mirror, d. 148 - 10.000 Hz shot).}
\label{fig:fig2}
\end{figure}

\begin{figure}[tbp]
\begin{center}
\epsfig{file=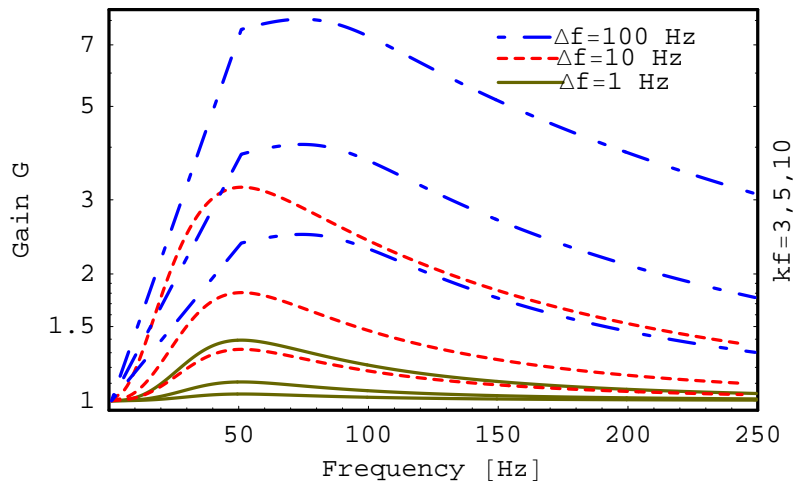}
\end{center}
\caption{Gain G in the frequency range 1-250 Hz for three levels of abatement of noise 
($ k_f = 3, 5, 10$), relative to an improvement in a bandwidth of 1, 10, 100 Hz.}
\label{fig:fig3}
\end{figure}

\begin{figure}[tbp]
\begin{center}
\epsfig{file=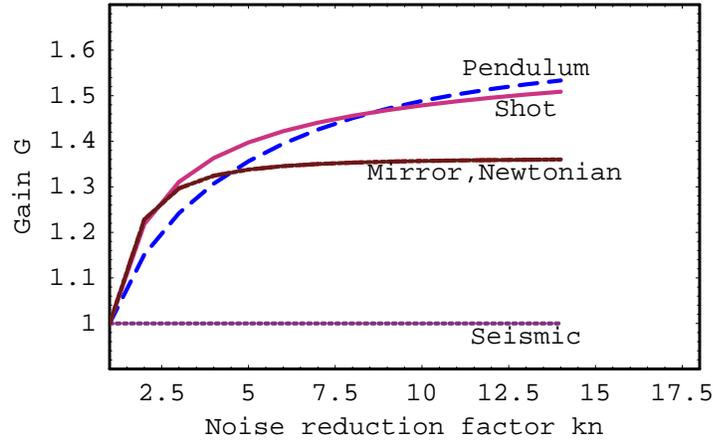}
\end{center}
\caption{Gain G versus abatement of noise $k_n$ for the four major noises (Newtonian, thermal pendulum, thermal mirror, 
shot).}
\label{fig:fig4}
\end{figure}

\begin{figure}[tbp]
\begin{center}
\epsfig{file=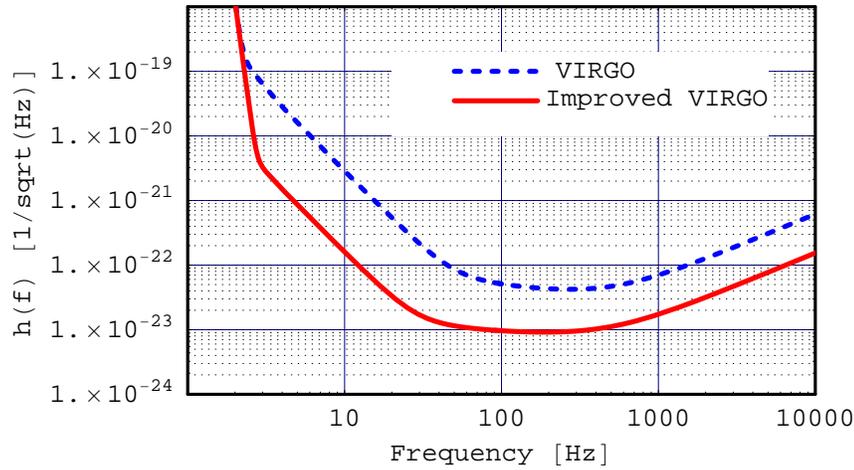}
\end{center}
\caption{VIRGO strain sensitivities $h(f)/\sqrt{Hz}$: the actual configuration and an example of an improved configuration (Punturo, 2004).}
\label{fig:fig5}
\end{figure}

\begin{figure}[tbp]
\begin{center}
\epsfig{file=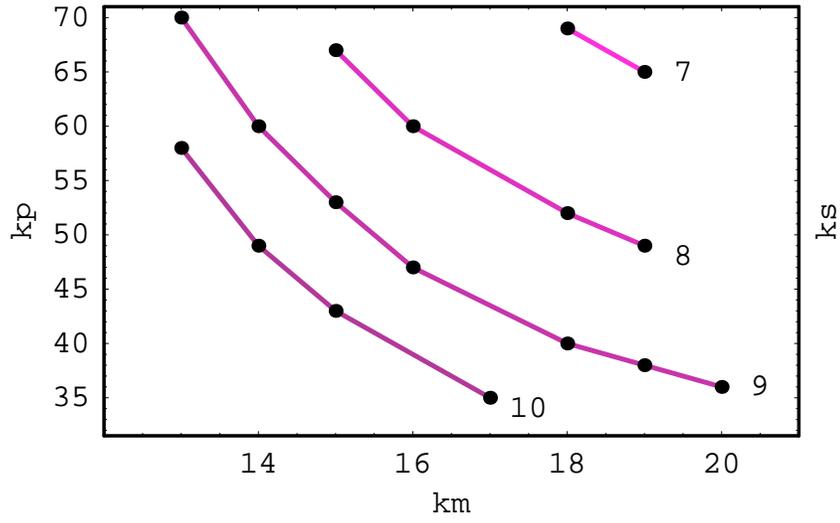}
\end{center}
\caption{The dots correspond to the integer values of the $k_n$ reduction factors for thermal pendulum and mirror noises, parametrised by the shot noise. Any threefold combination leads to an equal 
gain G = 16 for VIRGO and therefore a bi-monthly detection rate for NS-NS coalescences, as advanced LIGO.}
\label{fig:fig6}
\end{figure}

\end{document}